# Optical probing of fractal and multifractal connection to structural disorder in weakly optical disordered media: Application to cancer detection

SANTANU MAITY, MOUSA ALRUBAYAN, ISHMAEL APACHIGWAO, DHRUVIL SOLANKI, AND PRABHAKAR PRADHAN,*

*Department of Physics and Astronomy, Mississippi State University, Mississippi State, MS 39762.*

*Corresponding Authors:  PP

**Abstract:** The light scattering experiment establishes a relationship between refractive index fluctuations and fractal dimension in weakly scattering tissue-like media. Based on the box-counting approach, an analytical model is developed and shows that the fractal dimension has a functional dependency on the structural disorder or refractive index fluctuation for short-range correlation and approximately linearly depends on each other for tissue-like media. Several parametric imaging systems can be connected using this approach. Further, tissue's weak multifractality optical scattering is explored using the box-counting method. It is shown that with a functional transformation, the distribution follows lognormal distributions.

## 1. Introduction:

Fractals are self-similar objects; a disordered system with fractal properties is measured by fractal dimension. Tissues/cells are weakly optical disordered fractal systems. [1–5]. It is now known that fractal dimension changes with the progress of several diseases, including cancer and brain abnormalities [6–8] . A deep connection between the change in fractal dimension and structural disorder is challenging [9]. This paper applies the box-counting method to establish a relationship between structural disorder and fractal dimension for short-range correlation refractive index fluctuations. An analytical model is introduced, which shows that the fractal dimension has a functional dependency on refractive index fluctuation and a linear dependency on weakly disordered media, such as cells/tissues. Using this connection, one can gain insight into the different imaging systems that image the structural disorder-related scattering mean free path or refractive index fluctuations. Tissue/cells also have a multifractal nature. Although there are several approaches to quantifying the multifractality of a sample, accurate quantification is challenging. We introduced a functional transformation of the fractal dimension at each point of fractal dimenssion, and the results show a good Gaussian distribution that works well for the different cancer tissue sample cases we studied. This paper mainly addresses weakly disordered media targeting tissue-like samples and refractive refractive index media probed by optical transmission microscopy

## 2. Fractal dimensions of weakly disordered media: Methods, Techniques, and Results

*2.1. Box-counting method for fractal dimension in biological cells/tissue using optical transmission microscopy.* A structure's fractal dimension is essentially a measure of how self-similar the structure is; the higher the dimension, the more self-similar it is, with more filling or less porosity[10]. Box-counting is one of the straightforward methods to calculate fractal dimensions with optical transmission microscopy. This method works by placing the fractal structure on an evenly spaced grid and counting the boxes needed to cover the structure [11,12]. It follows the Euclidean definition of fractal dimension: $N_i(r_i) \times (r_i)^{Df} = N_j \times (r_j)^{Df}$ = Constant, where $N_i(r_i)$ is the number of boxes at length scale $r_i$. The average fractal dimension $D_f$ is then calculated as an ensemble-averaged slope of $\ln(1/r)$ vs. $\ln(N(r))$ curve, with varying length scales $r$. Fractal dimension,

$$D_f = \ln(N(r))/\ln(1/r) \qquad (1)$$

*2.3. Model to calculate the relationship between structural disorder and fractal dimensions:* Two main types of scattering arise from the interaction of light with cells and tissues. One is bulk scattering due to the contrast with the environment, and the other is scattering from the spatial RI (refractive index) fluctuations. Considering a lattice model with the fractal random cuts, our previous study showed that the random cuts and real fractal generating models like diffusion limited aggregation have properties similar to those of tissue-like refractive index media.

One can derive the standard deviation of the random cuts fractal model in a short-range correlation [9]. Consider that the sample's integer dimension (D) will be randomly cut, and $N_0$ is the total number of filled pixels. At a particular instant in a partially field fractal lattice, $N_c$ number of pixels are cut, for the square of the standard deviation is $\sigma^2$,

$$\sigma^2 = \sum[n(i) - <n>]^2/N_0, \qquad (2)$$

where $n(i)$ is the number of box entries at the point, either 0 or 1 ( binary). n=<n(i)> is the mean of n(i), or $<n(i)> = (N_0 - N_c)/N_0$, $N_0$ is the total number of lattice points. From (1) and (2), one can get the relationship between the structural disorder parameter $\sigma^2$ and $D_f$ and $N_0$,

$$\sigma^2 = N_0^{(\frac{D_f}{n}-1)} - N_0^{(\frac{2D_f}{n}-2)}, \tag{3}$$

The inverse equation connecting the fractal dimension with the structural disorder σ²,

$$D_f = n \times \ln\left(\frac{N_0 + N_0\sqrt{1-4\sigma^2}}{2}\right)/\ln(N_0). \tag{4}$$

In the early stages of tissue malignancy, the mass density fluctuations change without changing the mean due to the rearrangement of the mass density. Calculating the change of standard deviation with the fractal dimension shows a functional dependence for all three dimensions, 1D, 2D, and 3D, as shown in the plot in Fig. 1 using equation Eqns. (3) and (4).

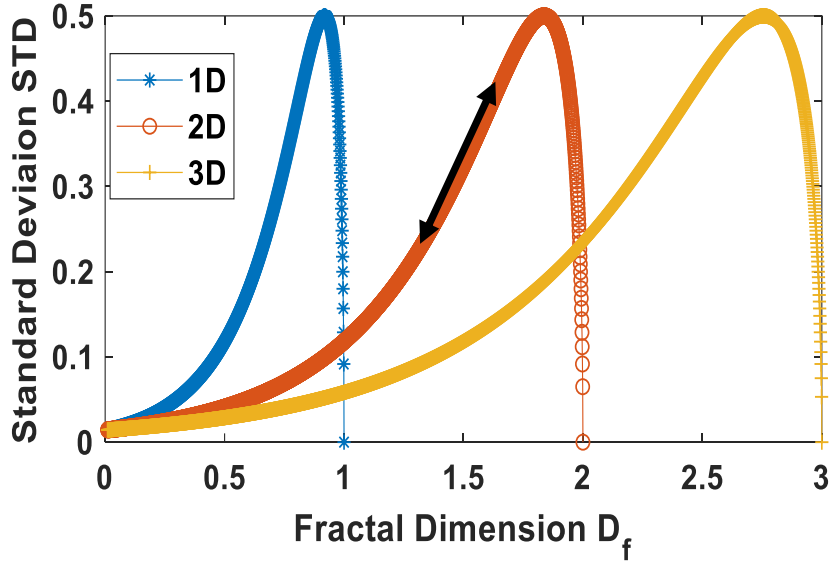

**Fig.1. A.** Variation of the standard deviation of the fractal sample σ with the fractal dimension by plotting Eqs. (3)-(4). The functional dependency's shape peaks near to $D_{fmax}$ (=1, 2, and 3) for each dimension. The main region of interest for the 2D tissue sample is around $D_f$ =1.5 to 1.8, where $D_f$ and σ are linearly dependent, as shown in the black-arrow line.

*2.2. Dependence of fractal dimension value $D_f$ and structural disorder $L_d$ values for major cancer cases:*
We use thin tissue samples to image in reflection/transmission modes in most optical microscopy. The fractal dimension of these tissue samples of various organs generally ranges from Df ~1.5 to 1.8. As shown in Fig. 2, in this range of 2D samples, the functional dependency of refractive index fluctuation/standard deviation and fractal dimension depends linearly, as shown in Fig. 2.

## 2.3. Large-volume Optical Transmission Microscopy Imaging and box-counting method of $D_f$ calculation.

*Tissue samples:* Our study used tissue microarray (TMA) samples of different cancers. TMA samples provided around 5 microns thick samples of control and multiple stages of cancer in a single slide. We analyze the following control and cancer (Stage I, II, and III) tissues: breast, colon, prostate, and pancreatic.

*Tissue imaging:* Details of the microscopy system are reported somewhere else. In brief, the Olympus BX61 microscope, CCD camera, and PRIOR Test Control (OptiScan software) were used for large-volume bright-field transmission imaging. Rather than the traditional way of manual imaging, we used the auto-scan feature of the microscope. To image the tissues, we used a conventional Olympus BX61 motorized system microscope with a 40x objective (UIS2) series (or higher), and a CCD camera mounted on the top of a BX61 microscope. In addition, we used the PRIOR Test Control via OptiScan software to move the microscope stage to synchronize with the camera while taking the scattering micrographs in the transmission mode. We will also add the autofocus option for a better-focused image performed by using the GUI Interface. The most significant squares within each observation site for a sample (~1.5 mm) were determined and divided into numbers proportional to the objective lens used (40x). Altogether, ~ 200 tissue microscopic micrographs were generated from each sample type for analysis.

We study breast, colon, pancreatic, and prostate cancers to determine the functional relationship between the refractive index fluctuations $\sigma^2(dn)$ and the fractal dimension $D_f$ for 2D thin samples on a glass TMA slide. First, tissue was imaged in the microscope's transmission mode. Then, standard binary approaches were used to transfer 2D micrographs to binary pictures, and then the box-counting algorithms, the ensemble-averaged fractal dimension of each cancer case were calculated for control, Stage I, Stage II, and Stage II[7]. Results of 4 major cancers, breast, colon, prostate, and pancreatic, cancers are shown here for the dependency of the fractal dimension and refractive index fluctuations. The fractal dimensions were calculated from the transmission optical microscopy micrographs and box-counting method, while the refractive index fluctuations shown in the pictures are calculated using partial wave spectroscopy (PWS) experiments that measure $Ld \sim dn^2$ for a short-range correlation, discussed in detail somewhere else [13,14].

The experimental results show a linear dependence between the fractal dimension and the cancer progression: control, Stage I, Stage II, and Stage III, as shown in Fig.2. $D_f$ is approximately limited to a value ~ of 1.5 to 1.8. This is consistent with the analytical calculation in the 2D region of Fig.1 for the $D_f$ value range of 1.5 to 1.8.

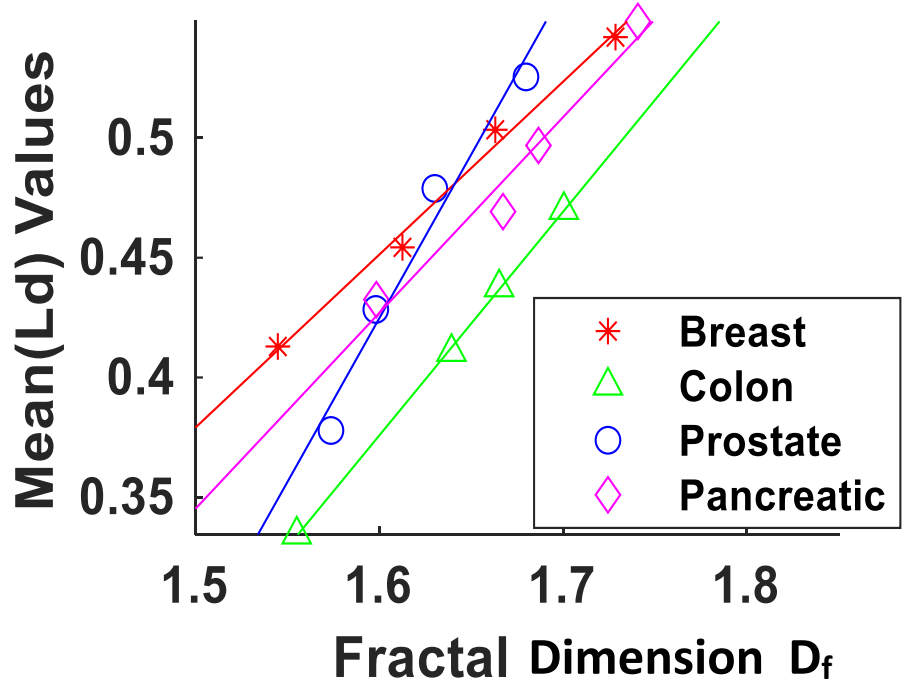

**Fig.2. A.** *Breast, Colon, Prostate, and Pancreatic cancer tissues $D_f$ vs σ.* Variation of fractal dimension with the different stages of cancer. The plot show as a linear dependence of RI fluctuations and fractal dimension. For each cancer case, the bottom of the line markers are for control and increasing values up makers are Stage I, Stage II, Stage III.

*Imaging in different parameter spaces*, *mimicking different modalities:* Based on PWS images of tissue samples $L_d \sim \sigma^2 \sim dn^2$ image, one can reproject to $D_f$ image and vice versa, using Eqs. (7) and (8). As described above, there is a functional relationship between fractal dimension and the structural disorder in a short-range correlation of sample, We performed a 2D imaging for $L_d$ on the same sample: Fractal dimension analyses and standard deviation.

Images of $L_d \sim \sigma 2$ by PWS and Df imaging by optical microscopy were performed on the same tissue samples. As can be seen, we were able to reproduce similar images. This is because of the region's linear dependency in 2D ( middle plot). The key concept is here to reduce several modal imaging from a simple transmission microscope and box-counting algorithm of fractal dimension.

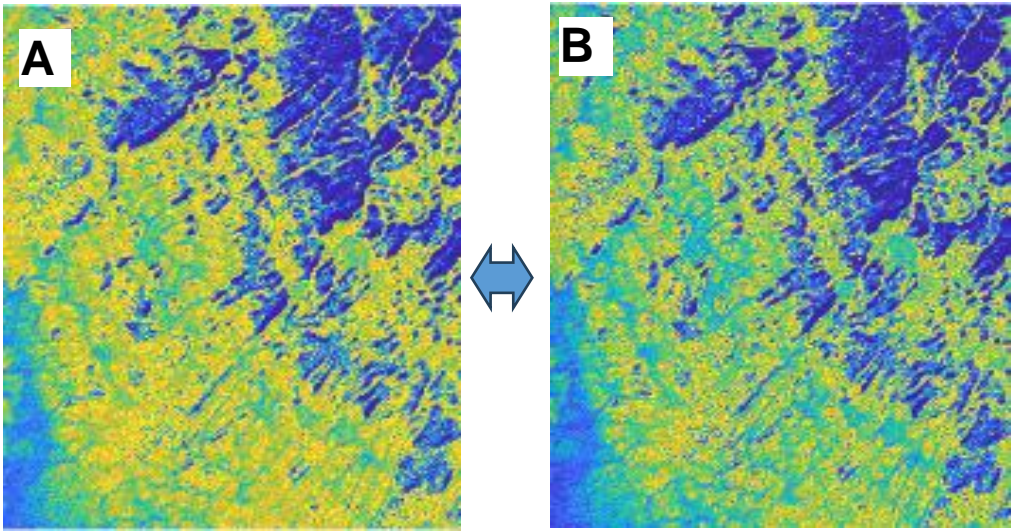

**Fig.3.** Transformation of **(A)** $D_f$ image of an optical microscope to **(B)** PWS Ld~$\sigma^2$~$dn^2$ image of PWS and vice versa. They look similar within a multiplication factor.

## 3. Multifractality of the tissue samples: Techniques and Results

Tissues are heterogeneous self-similar media and can be well-parametrized by a mono fractal. However, it has some signature of multifractality. We perform the spectrum analyses or $f(\alpha)$ vs. α test of pancreatic, colon, and lung cancers. This is the main test via box-counting methods to get the signature of the multifractality of a sample. It can be noted that control tissues are slightly multifractal, and the fractality changes with the progress of carcinogenic stages, Stage I , Stage II, and Stage III. [41-43]

*Multifractality Analyses* [15,16]: With some tricks, one can plot the multi-fractality of 2D thin tissue samples using the same box-counting method. This is a relatively easier method to check the multifractality spectra of the tissue. A quantitave value is challending, it shoes the change of the lattice with with probabilistic power changes of the lattice intensity. In this case, one first divide the whole intensity of the $L \times L$ sample size 2D samples into a certain length scale $\varepsilon \times \varepsilon$ boxes, then calculate the probability of pixels/mass present at an *i*th cell: $P_{\varepsilon,i} = N_\varepsilon(i)/N_{total}$ ~ $e^{\alpha i}$, where $N_\varepsilon(i)$ is the number of pixels at *i*th box in length scale $\varepsilon$ then $\mu_{i(Q,\varepsilon)} = P^Q_{i(Q,\varepsilon)}/\sum_{i=1:n\varepsilon}(P^Q_{i(Q,\varepsilon)})$, we will plot the spectral function:[15]

$$f(\alpha_Q) = Q \times \alpha_Q - \tau_Q = \sum_{i=1}^{n\varepsilon} \mu_{i(Q\varepsilon)} \times \ln(\mu_{i(Q\varepsilon)}) \qquad (5)$$

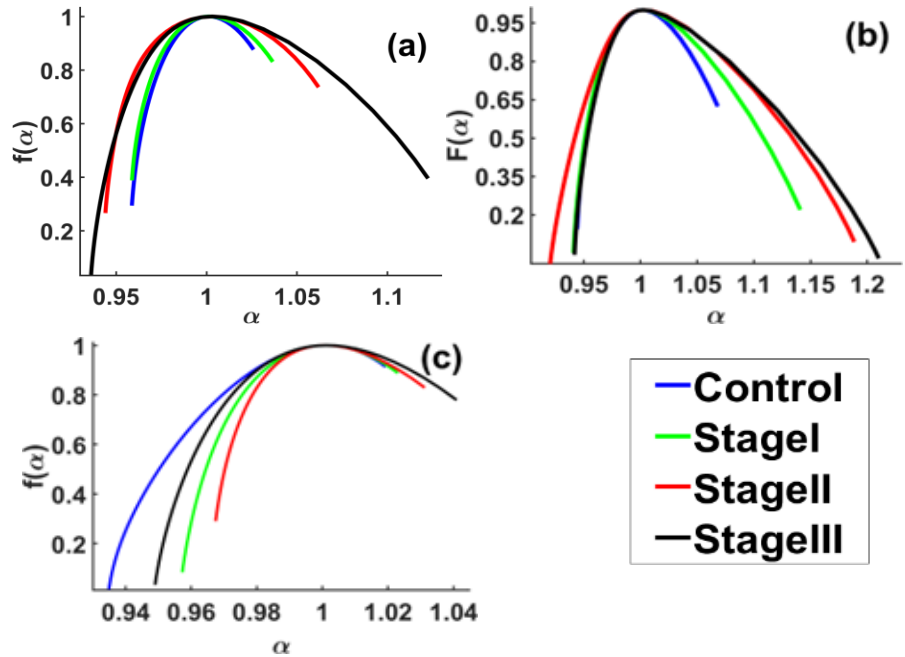

**Fig. 4.** Multifractal spectrum *α vs. f(α)* plots for (a) Pancreatic, (b) colon, and (c) lung for controls and Stage I, Stage II, and Stage III tissue samples of each. (N=10.). Figure shows small changes in the spectrum.

where the increased power value of power Q is scanned over a range of -10 to +10, and $n_\varepsilon =(L\times L)/(\varepsilon\times\varepsilon)$. This provides a multifractality test to determine how the different parts of the tissues are filled and fluctuate in density. The variation is small for mono-fractals and higher for multifractals. With multifractal tests, it has a broader range [15-16]. In Fig. 4, we plotted the *f(α) vs α* plot to test the multifractality. It can be seen from the figure that the control has some multifractality, and there is a small spread with the progress of cancer. This algorithm does not provide good quantification but a visual inspection of the range relative to the control. This holds for all types of tissues that are within a weak disordered media. Multifractal holds good for relatively sparse media in density distribution; however, tissues are mostly heterogeneous disordered media and have small sparsity.

## 4. A new approach of multifractal quantification for weakly disordered media

Most tissue samples belong to weakly disordered media, which are multifractals. The $D_f$ distribution (non-binary) is non-Gaussian-like, with a slightly extended tail at one side. The next step for the non-Gaussian distribution with a tail is to check for the log-normal distribution of $D_f$, that is, test for the P(ln(Df)) distribution as Gaussian. In Fig. 5, P(Df) and P(ln(Df)) plots are shown. P(Df) shows a tailed distribution. P(ln(Df)) does not show either Gaussian distribution.

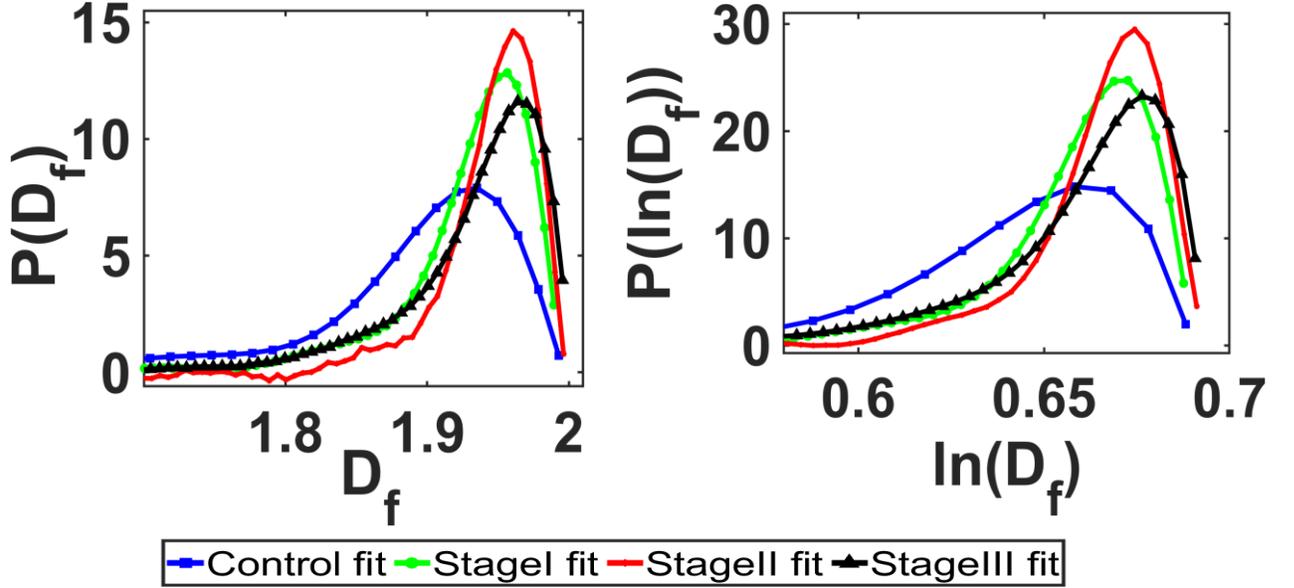

**Fig. 5.** (a) The fractal *$D_f$* vs. *P(Df) plot shows a slightly extended-tailed* distribution. (b) The plots of *ln($D_f$) vs. P(ln($D_f$))* show a more Gaussian distribution.

*A new functional approach to quantify the weak multifractal.*

In condensed matter physics, it is known that a variable transformation can extend a closed parameter to capture the larger space. (which is called the Laudauer formula). Applying a similar concept, one can form a large tail distribution from a small-tailed distribution or make a more lognormal distribution such that the ln of the new variable distribution will be of t a lognormal distribution.

We define the new $D_{tf} = f(D_f)$ that can unfold the $D_f$ in a new functional space :

$$D_{tf} = \frac{D_f}{D_{fmax} - D_f} \qquad (6)$$

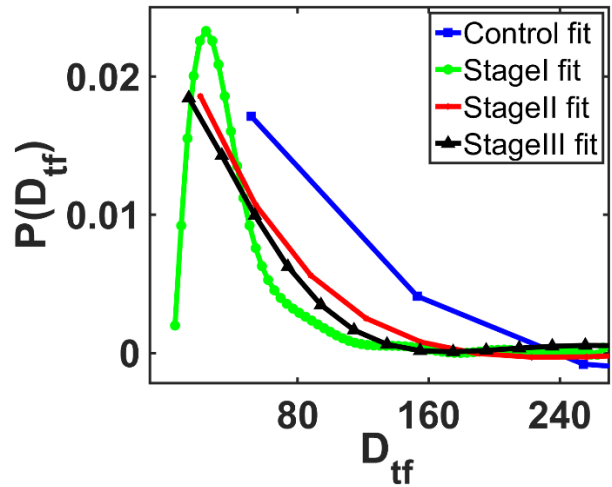

**Fig. 6.** P($D_{ft}$) vs $D_{ft}$ plot shows a lognormal type ditribution with a long tail, one expects the ln(Dft) will be be normal distribution.

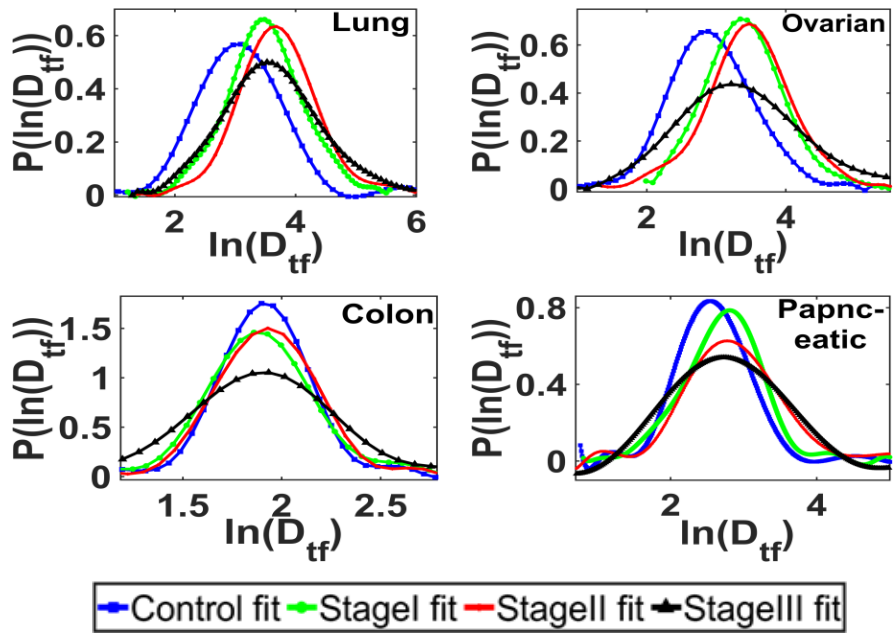

**Fig. 7.** Best polynomial fits for lung, ovarian, colon, and pancreatic cancers. The P(ln($D_{ft}$)) vs $D_{ft}$ plot shows a Gaussian-like distribution in the best polynomial fits.

where $D_{fmax}$ is the maximum fractal dimension in dimensions 1, 2, 3, or Dimension, that is 1, 2, and 3. This function has a broader range, with a very long tail, which is a more lognormal distribution. The ultimate aim is to make ln($D_{tf}$) function as a Gaussian distribution.

To verify our extended phenological transform function P(ln(Dft) distribution, we demonstrate the ln(Dft) vs. P(ln(Dft)) function in the above Fig. 7. It can be seen that most of the functions follow an approximately

Gaussian distribution, for the major cancer cases: lung, ovarian, colon, and pancreatic and the advantage and the standard deviation increase with the progress of cancer from control to Stage I, Stage II, and Stage III.

In Fig. 8, we plot Gaussian fittings of the above distribution in Fig.(7), which looks good fit with Gaussian.

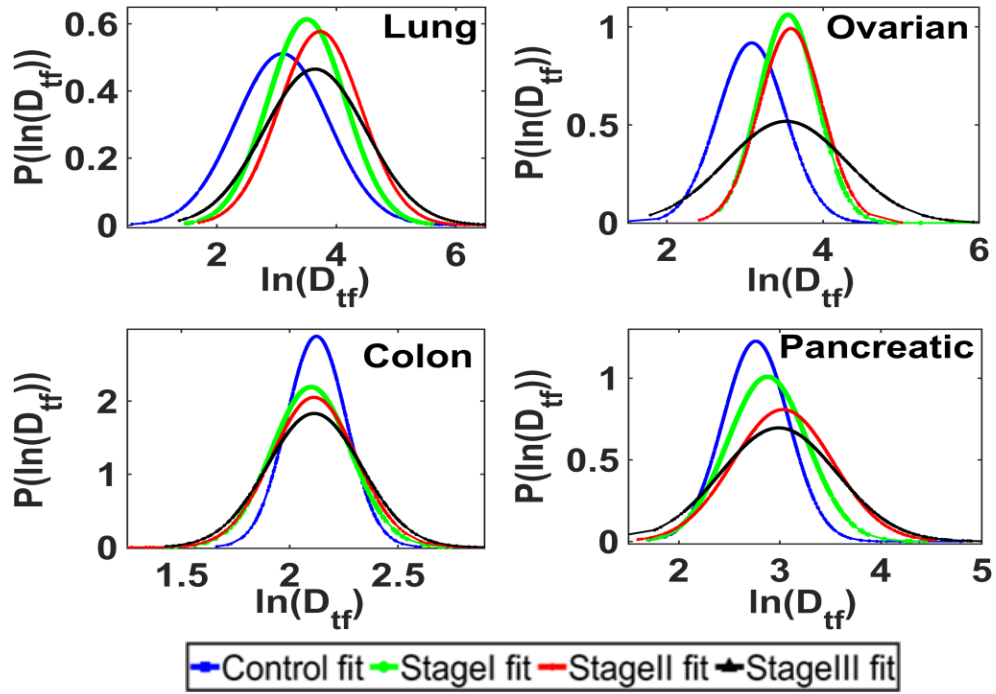

**Fig.8.** Best Gaussian fittings of Fig. (7) for lung, ovarian, colon, and pancreatic cancers, $P(\ln(D_{ft}))$ vs $D_{ft}$ plots show a good fit to Gaussian function where the mean/STD of $P(\ln(D_{ft}))$ changes with the progress of cancer from control to different stages.

## 5. Conclusions:

This paper developed a relationship between fractal dimension and structural disorder or refractive index fluctuations. An analytical model developed based on the box-counting method shows that the fractal dimension has a functional dependency on the structural disorder or refractive index fluctuation for short-range correlation and approximately linearly depends on each other for tissue-like media. Furthermore, it was shown that several parametric imaging systems can be connected using this approach. Finally, tissue's weak multifractality optical scattering is explored using the box-counting method using a new functional transformation and shown that the new variable distribution follows lognormal distributions.